\documentclass[runningheads]{style/llncs}

\usepackage[T1]{fontenc}
\usepackage{lmodern} 
\usepackage[varqu]{inconsolata} 
\usepackage{bbding} 
\usepackage{booktabs}
\usepackage{listings}
\usepackage{graphicx} 
\usepackage{booktabs} 
\usepackage{xcolor}
\usepackage{hyperref}
\usepackage{placeins}

\lstdefinestyle{simplehmstyle}{
    basicstyle=\footnotesize\ttfamily,
    breaklines=true,             
    showstringspaces=false,      
    tabsize=2,                   
    frame=single,
    keywordstyle=\bfseries,
    morekeywords={
        String, Int, Float, Data,
        Message, Argument,
        DebateRequest, RetrievalResponse, SystemResponse,
        UserTurn, Simulation,
        EvaluationRequest, EvaluationResponse,
    },
    commentstyle=\color{green!50!black}, 
    morecomment=[l]//,                   
    morecomment=[l]\#,                   
}
\begin{document}
\title{DS@GT ARC at Touch\'e: Large Language Models for Retrieval-Augmented Debate}
\titlerunning{DS@GT ARC at Touch\'e}

\author{
    Anthony Miyaguchi\textsuperscript{(\Envelope)}\orcidID{0000-0002-9165-8718} 
    \and 
    Conor Johnston\orcidID{0009-0001-1777-9255}
}

\authorrunning{A. Miyaguchi et al.}

\institute{Georgia Institute of Technology, North Ave NW, Atlanta, GA 30332, USA \\
\email{\{acmiyaguchi,cjohnston\}@gatech.edu}}

\maketitle

\begin{abstract}
    We extend the DS@GT ARC working-note submission to the Touch\'e 2025 Retrieval-Augmented Debate task.
    The task has two subtasks: generating the next utterance in a simulated debate, and evaluating debate responses according to the Gricean maxims of Quantity, Quality, Relation, and Manner.
    The DS@GT ARC submission consisted of six leading LLMs from three providers through a retrieval-augmented prompting pipeline.
    We summarize the results from the working paper and explore whether multi-LLM evaluator agreement is a reliable proxy for official evaluation performance.
    The analysis shows that frontier LLM systems are strong response generators, and as evaluators they agree strongly within model families.
    However this consensus does not reliably track the official evaluation target, with the largest gap on the Quality maxim.
    The accompanying source code for this paper is located at \url{https://github.com/dsgt-arc/touche-2025-rad} and \url{https://github.com/dsgt-arc/touche-2025-rad-analysis}.

\keywords{Large Language Models (LLM) \and Retrieval-Augmented Generation (RAG)
  \and Argumentative Systems \and Self-Evaluation \and Conversational AI
  \and DS@GT}
\end{abstract}

\section{Introduction}

The DS@GT team participated in the Retrieval-Augmented Debating (RAD) task as part of the Touché 2025 competition \cite{kiesel:2025}.
The RAD task evaluates argumentative systems in a controlled turn-based simulation where the system chooses to defend its previous response, or attack the opponent's.
In the first subtask, the system receives the current debate state, retrieves candidate supporting arguments, and then generates an utterance.
In the second subtask, the system evaluates debate responses according to the Gricean maxims chosen by the organizers: Quantity, Quality, Relation, and Manner \cite{grice1975logic}.

We describe our approach which leverages retrieval-augmented generation (RAG) using the ClaimRev corpus \cite{skitalinskaya2021claimrev} and large language models (LLMs).
This paper extends the original DS@GT working note \cite{miyaguchi2025dsgt}.
Sections~\ref{sec:methodology} and~\ref{sec:results} summarize the submitted systems and their official evaluation, alongside new post-hoc analyses of the organizer-released evaluation data.
We interpret the relationship between evaluator consensus and official evaluation scores in Section~\ref{sec:discussion}.

We design an evaluation pipeline to assess our responses across the required dimensions.
We hypothesize frontier LLMs can produce strong retrieval-augmented debate responses in Touché RAD 2025 and analyze the reliability of the evaluation process itself.
We organize the follow-up analysis around three questions: how reliable multi-LLM judge consensus is, whether weak supervision can improve evaluator calibration, and whether evaluator consensus aligns with official evaluation scores.

\section{Related Work}

LLMs are effective debaters \cite{khan2024debatingpersuasivellmsleads}.
Although not examined in a debate environment, LLMs enhanced with RAG systems have been shown to improve in accuracy and consistency in responses \cite{hagstrom2023effectscalingretrievalaugmentation}.  
These systems are well-suited for debate because they require strong argumentative skills, along with the ability to recall factual knowledge. 
Studies have shown that LLMs can surpass human performance in debate evaluations \cite{liu2024empiricalanalysislargelanguage}.

The original DS@GT working note deployed six publicly available models across three providers for retrieval-augmented debate response generation and evaluation, finding that LLMs performed well with retrieved arguments but tended toward verbose responses and consistent evaluation behavior \cite{miyaguchi2025dsgt}.
The SINAI working note instead emphasized an open-source LLaMA3.1-8B-Instruct approach, using structured multi-step prompting for generation and zero-shot, few-shot, and analyzer-style prompting strategies for evaluation \cite{vallecillo-rodriguez:2025}.

\section{Methodology}
\label{sec:methodology}

We develop a debate system for both the response and evaluation RAD subtasks using large language models.
The response subtask generates an appropriate response to an ongoing debate.
The evaluation subtask assesses the quality of the given response in the context of the debate.
The interfaces adhere to the GenIRSim API, which is used to simulate and evaluate debates \cite{kiesel2024will}.

Each debate is defined by an underlying topic, simulated user, turn history, and the retrieved arguments for the responding system.
The official response generation test collection consists of 100 topics.
Each submitted run is simulated for five system turns, yielding 500 responses per run.
The official response data distributed by the organizers consists of eight systems: six of the DS@GT submitted systems, one SINAI submitted system, and the organizers' baseline.
Subtask 1 is scored as the proportion of responses fulfilling the Gricean maxims as per official binary annotations, with submissions ranked through an average of all the scores.
Subtask 2 is scored as a classification task with precision, recall, and F1 per maxim, ranked by an average of the F1 scores.
Post-competition the organizer annotations and evaluator scores were made available for analysis.
Figure~\ref{fig:example_debate} shows one exchange from the released corpus.

\begin{figure}[tbp]
    \centering
    \input{sections/listings/35_listing_example_debate}
    \caption{
        An example debate exchange from the organizer-released corpus (topic 7, first system turn of our gpt-4.1 run).
        The topic, simulated user turn, and retrieved arguments are provided by the organizers while the system generates the response.
        Official annotators marked all four maxims unsatisfied because the system argues the same stance as the user, while all six frontier LLM judges marked all four satisfied.
    }
    \label{fig:example_debate}
\end{figure}

\begin{table}[tbp]
    \centering
    \caption{Models used for RAD experiments via OpenRouter.}
    \label{tab:llm_models}
    \begin{tabular}{@{}l c r r r@{}}        
        \toprule
        \textbf{Model Name} & \textbf{Release Date} & \textbf{Context} & \textbf{Input} & \textbf{Output} \\
                            &                       & \textbf{(tokens)}     & \textbf{(\$/M)} & \textbf{(\$/M)} \\
        \midrule
        anthropic/claude\_opus-4 & 2025-05-22 & 200,000 & 15 & 75 \\
        anthropic/claude\_sonnet-4 & 2025-05-22 & 200,000 & 3 & 15 \\
        google/gemini-2.5-flash-preview-05-20 & 2025-05-20 & 1,048,576 & 0.15 & 0.60 \\
        google/gemini-2.5-pro-preview & 2025-05-07 & 1,048,576 & 1.25 & 10 \\
        openai/gpt-4.1 & 2025-04-14 & 1,047,576 & 2 & 8 \\
        openai/gpt-4o & 2024-05-13 & 128,000 & 2.50 & 10 \\
        \bottomrule
    \end{tabular}
\end{table}

\begin{figure}[tbp]
    \centering
    \input{sections/listings/31_listing_debate_response}
    \caption{
        The function signature of the responding endpoint the GenIRSim system expects for a valid simulation.
        The messages are passed in OpenAI-compatible chat completion format, with the roles of "user" and "assistant".
    }
    \label{fig:debate_interface}
\end{figure}

\begin{figure}[tbp]
    \centering
    \input{sections/listings/32_listing_debate_evaluation}
    \caption{
        The function interface for the evaluation portion of the GenIRSim API.
        A separate request is made for quantity, quality, manner, and relation.
    }
    \label{fig:evaluate_interface}
\end{figure}

\subsection{Debate Response}
\label{sec:debate-response}

The debate response is split into a retrieval and generation phase.
In the retrieval phase, we obtain the top ten documents from the argument search system as evidence for a response.
The search system utilizes organizer-hosted Elasticsearch and Stella embeddings to retrieve relevant debate documents \cite{zhang2024jasper}. 
The generation phase prompts an LLM, shown in Figure \ref{fig:prompt_debate}, with evidence and history of the debate.
We then respond to the debate request using the interface described in Figure \ref{fig:debate_interface}.
The debate topic, retrieved arguments, and prior turns are inputs, while the model output is the response returned to GenIRSim (Figure~\ref{fig:example_debate}).

We made six submissions to the response-generation task with models selected from OpenRouter, an LLM gateway service that provides unified access to a diverse set of providers.
We select flagship models with context windows ranging from 128k to 1m tokens listed in Table~\ref{tab:llm_models} across three providers in May and June 2025.
Date-suffixed identifiers pin an exact snapshot, while the remaining identifiers resolved to the provider's then-current version of the listed release.
Prices are reported in U.S. dollars per million input or output tokens.

In exploratory tests, we generated a set of six random topics from the ClaimRev corpus \cite{skitalinskaya2021claimrev}.
The simulated debate has a maximum of three turns per model, resulting in six simulations per model for a total of 18 utterances.
We fed each model its previous simulations to verify the pipeline end-to-end before the official submissions.

\begin{figure}[tbp]
    \centering
    \input{sections/listings/33_listing_prompt_debate}
    \caption{
        The prompt is used to generate a debate response in the first subtask.
        Evidence from the claims database is formatted as YAML at the start of the prompt, while the context of the entire argument is serialized at the end of the prompt.
    }
    \label{fig:prompt_debate}
\end{figure}

\begin{figure}[tbp]
    \centering
    \input{sections/listings/34_listing_prompt_evaluation}
    \caption{
        The prompt is used to generate all measures of the requested debate.
        The call to the LLM leverages structured output through the chat completion API provided by OpenRouter.
    }
    \label{fig:prompt_evaluation}
\end{figure}

\subsection{Debate Evaluation}
\label{sec:debate-evaluation}

We evaluate debate responses in a zero-shot manner and made six submissions using the LLM models listed in Table~\ref{tab:llm_models}.
We prompt the LLM to judge the most recent responses and the topic of the debate as per Figure \ref{fig:prompt_evaluation} with structured generation to generate numeric scores for quantity, quality, manner, and relation.
The web interface must expose four separate endpoints for the evaluation measures.
To limit the total number of LLM calls, we evaluate all four measures in a single call and memoize the result.

We ran toy simulation topics through the baseline simulation image provided by the committee.
The toy simulation consists of two topics: "Television is bad" and "Television is good."
For every model, we evaluated the baseline simulations of these debates (two topics, one debate per topic with three messages each) four times. 
We used the baseline image for the simulations to ensure a consistent set of debates for evaluating our results.

All runs were submitted through the TIRA platform \cite{froebe:2023}.
Because API keys cannot be exposed safely inside the TIRA environment, the submitted software is a thin proxy client that forwards requests to a self-hosted service, which handles logging and the LLM calls via OpenRouter.


\subsection{Evaluator Consensus and Calibration}
After the official leaderboard period elapsed, we shifted our focus to the analysis of the model behavior, in particular behavior of LLMs as judges.
The eight systems are the six DS@GT response-generation runs listed in Table~\ref{tab:llm_models}, the organizer baseline, and the SINAI/Lewis-carroll response-generation run.
Each system is evaluated on 100 topics with five system turns per topic, producing 4,000 response turns joined to 49,600 organizer-released annotation/evaluation rows with no missing response joins.
We look at evaluator consensus through pairwise Pearson correlation computed over response turns separately per maxim and mean averaged.

Given the evaluator agreement, we tested whether a weak-supervision ensemble could improve over simple multi-LLM averaging.
Our approach follows FlyingSquid-style label modeling \cite{pmlr-v119-fu20a} and is motivated by weak-verifier aggregation work (Weaver) \cite{saadfalcon2025shrinking}.
The voters are the six frontier LLM judges and ten surface heuristics covering length, lexical overlap, retrieval, readability, and repetition.
The expanded pool adds the seven SINAI LLaMA-8B judge runs as additional voters.
Each voter is binarized at its median and treated as a noisy vote over each Gricean dimension.
The label model then estimates latent voter accuracies and calibrated system scores.
This analysis is treated as a diagnostic for evaluator consensus.
Additionally, we tested a dependency-aware version of the label model, where correlated voters such as same-family LLMs and length-related heuristics were explicitly linked.

Since the weak-supervision experiments estimate which judges align with the latent consensus, we must evaluate whether that consensus corresponds to the official evaluation target.
To test this, we compared latent judge accuracy estimates against the official Subtask 2 F1 scores.
Official F1 measures performance against the lab organizers' evaluation target, while latent accuracy measures agreement with the weak-supervision model's inferred consensus, which needs to be taken into consideration during analysis.
\section{Results}
\label{sec:results}

\subsection{Leaderboard}

Table~\ref{tab:model_evaluation_overall_avg} shows that the internal evaluator runs were generally strict, with Manner receiving the highest scores and Relation, Quantity, and Quality receiving lower scores.
In the internal simulation evaluations, the Anthropic models were the most strict for overall score, with Opus 4 and Sonnet 4 coming in at 0.2773 and 0.2681, respectively.
The standard deviations of both models were less than 0.2600, which is notably more consistent than those of the other models.
Google's Gemini 2.5 Flash Preview gave the highest overall score of 0.4068, while the Pro Preview gave an overall evaluation of 0.3527. 
OpenAI's GPT-4o gave an overall average of 0.3870, and GPT-4.1 was at 0.3170. 
These averages are low overall.
The models agree that Relation, Quantity, and Quality of the simulation responses were poor.
Manner was the only maxim rated favorably across all models.

\begin{table}[t]
\caption{Internal model evaluations across Gricean metrics, reported as
mean $\pm$ standard deviation, with the overall average computed from the
four metric means.}
\label{tab:model_evaluation_overall_avg}
\centering
\scriptsize
\setlength{\tabcolsep}{2.8pt}

\begin{tabular}{@{}lccccc@{}}
\toprule
\textbf{Model}
& \textbf{Overall}
& \textbf{Quantity}
& \textbf{Quality}
& \textbf{Relation}
& \textbf{Manner} \\
\midrule

google/gemini-2.5-flash
& \textbf{0.407}
& 0.315 $\pm$ 0.184
& 0.350 $\pm$ 0.220
& 0.269 $\pm$ 0.314
& 0.694 $\pm$ 0.148 \\

openai/gpt-4o
& 0.387
& \textbf{0.319 $\pm$ 0.076}
& \textbf{0.363 $\pm$ 0.092}
& \textbf{0.302 $\pm$ 0.194}
& 0.565 $\pm$ 0.127 \\

google/gemini-2.5-pro
& 0.353
& 0.167 $\pm$ 0.118
& 0.215 $\pm$ 0.212
& 0.223 $\pm$ 0.315
& \textbf{0.806 $\pm$ 0.139} \\

openai/gpt-4.1
& 0.317
& 0.219 $\pm$ 0.099
& 0.247 $\pm$ 0.102
& 0.167 $\pm$ 0.178
& 0.635 $\pm$ 0.093 \\

anthropic/claude\_opus-4
& 0.277
& 0.171 $\pm$ 0.059
& 0.243 $\pm$ 0.085
& 0.177 $\pm$ 0.254
& 0.519 $\pm$ 0.153 \\

anthropic/claude\_sonnet-4
& 0.268
& 0.215 $\pm$ 0.048
& 0.238 $\pm$ 0.079
& 0.120 $\pm$ 0.180
& 0.500 $\pm$ 0.186 \\

\bottomrule
\end{tabular}
\end{table}


We submitted multiple runs to the official Touché 2025 shared task leaderboard. 
The results for both subtasks are shown in Table~\ref{tab:touche25_subtask1_results} and Table~\ref{tab:touche25_subtask2_results}.
Our GPT-4.1 and Gemini-2.5 runs ranked among the top submissions in Subtask 1 and showed competitive results in Subtask 2.


\begin{table}[t]
\caption{Official Touché 2025 results for Subtask 1: proportion of
responses fulfilling each Gricean maxim across submitted runs (SINAI runs omitted)}
\label{tab:touche25_subtask1_results}
\centering
\small
\setlength{\tabcolsep}{4pt}

\begin{tabular}{@{}llccccc@{}}
\toprule
\textbf{Team}
& \textbf{Run}
& \textbf{Average}
& \textbf{Quantity}
& \textbf{Quality}
& \textbf{Relation}
& \textbf{Manner} \\
\midrule

Baseline
& baseline
& 0.62
& 0.35
& \textbf{1.00}
& 0.32
& 0.80 \\

DS@GT
& gpt-4.1
& \textbf{0.70}
& \textbf{0.95}
& 0.17
& 0.82
& \textbf{0.84} \\

DS@GT
& gemini-2.5-pro
& 0.65
& 0.94
& 0.26
& 0.74
& 0.67 \\


DS@GT
& gemini-2.5-flash
& 0.50
& 0.70
& 0.07
& 0.80
& 0.41 \\

DS@GT
& claude-opus-4
& 0.42
& 0.41
& 0.31
& 0.87
& 0.09 \\

DS@GT
& gpt-4o
& 0.42
& 0.20
& 0.02
& 0.86
& 0.58 \\

DS@GT
& claude-sonnet-4
& 0.38
& 0.35
& 0.05
& \textbf{0.94}
& 0.17 \\

\bottomrule
\end{tabular}
\end{table}

\begin{table}[t]
\caption{Official Touché 2025 results for Subtask 2: precision (P),
recall (R), and F1-score for each maxim classification task (SINAI runs omitted).}
\label{tab:touche25_subtask2_results}
\centering
\scriptsize
\setlength{\tabcolsep}{1.7pt}

\begin{tabular}{@{}llc*{4}{ccc}@{}}
\toprule
\textbf{Team}
& \textbf{Run}
& \textbf{Overall}
& \multicolumn{3}{c}{\textbf{Quantity}}
& \multicolumn{3}{c}{\textbf{Quality}}
& \multicolumn{3}{c}{\textbf{Relation}}
& \multicolumn{3}{c@{}}{\textbf{Manner}} \\

\cmidrule(lr){4-6}
\cmidrule(lr){7-9}
\cmidrule(lr){10-12}
\cmidrule(l){13-15}

& & \textbf{F1}
& \textbf{P} & \textbf{R} & \textbf{F1}
& \textbf{P} & \textbf{R} & \textbf{F1}
& \textbf{P} & \textbf{R} & \textbf{F1}
& \textbf{P} & \textbf{R} & \textbf{F1} \\

\midrule

Baseline
& 1-baseline
& \textbf{0.67}
& 0.57 & \textbf{1.00} & 0.73
& 0.24 & \textbf{1.00} & 0.38
& 0.78 & \textbf{1.00} & 0.87
& 0.52 & \textbf{1.00} & \textbf{0.68} \\

DS@GT
& gemini-2.5-flash
& 0.64
& 0.59 & 0.86 & 0.70
& 0.18 & 0.66 & 0.29
& 0.81 & 0.99 & 0.89
& 0.52 & 0.99 & \textbf{0.68} \\

DS@GT
& gpt-4o
& 0.64
& 0.59 & 0.88 & 0.71
& 0.17 & 0.63 & 0.27
& 0.82 & 0.99 & 0.89
& 0.52 & 0.97 & 0.67 \\

DS@GT
& gpt-4.1
& 0.62
& 0.58 & 0.75 & 0.65
& 0.15 & 0.52 & 0.24
& 0.82 & 0.98 & \textbf{0.90}
& 0.52 & 0.99 & \textbf{0.68} \\

DS@GT
& gemini-2.5-pro
& 0.62
& 0.59 & 0.67 & 0.63
& 0.17 & 0.52 & 0.25
& 0.84 & 0.97 & \textbf{0.90}
& 0.52 & 0.98 & \textbf{0.68} \\


DS@GT
& claude-sonnet-4
& 0.56
& 0.43 & 0.49 & 0.49
& 0.15 & 0.36 & 0.21
& 0.83 & 0.92 & 0.88
& 0.51 & 0.93 & 0.66 \\




DS@GT
& claude-opus-4
& 0.51
& 0.49 & 0.21 & 0.29
& 0.16 & 0.31 & 0.21
& 0.85 & 0.90 & 0.88
& 0.51 & 0.92 & 0.66 \\




\bottomrule
\end{tabular}
\end{table}



\begin{figure}[tb]
\centering
\includegraphics[width=\textwidth]{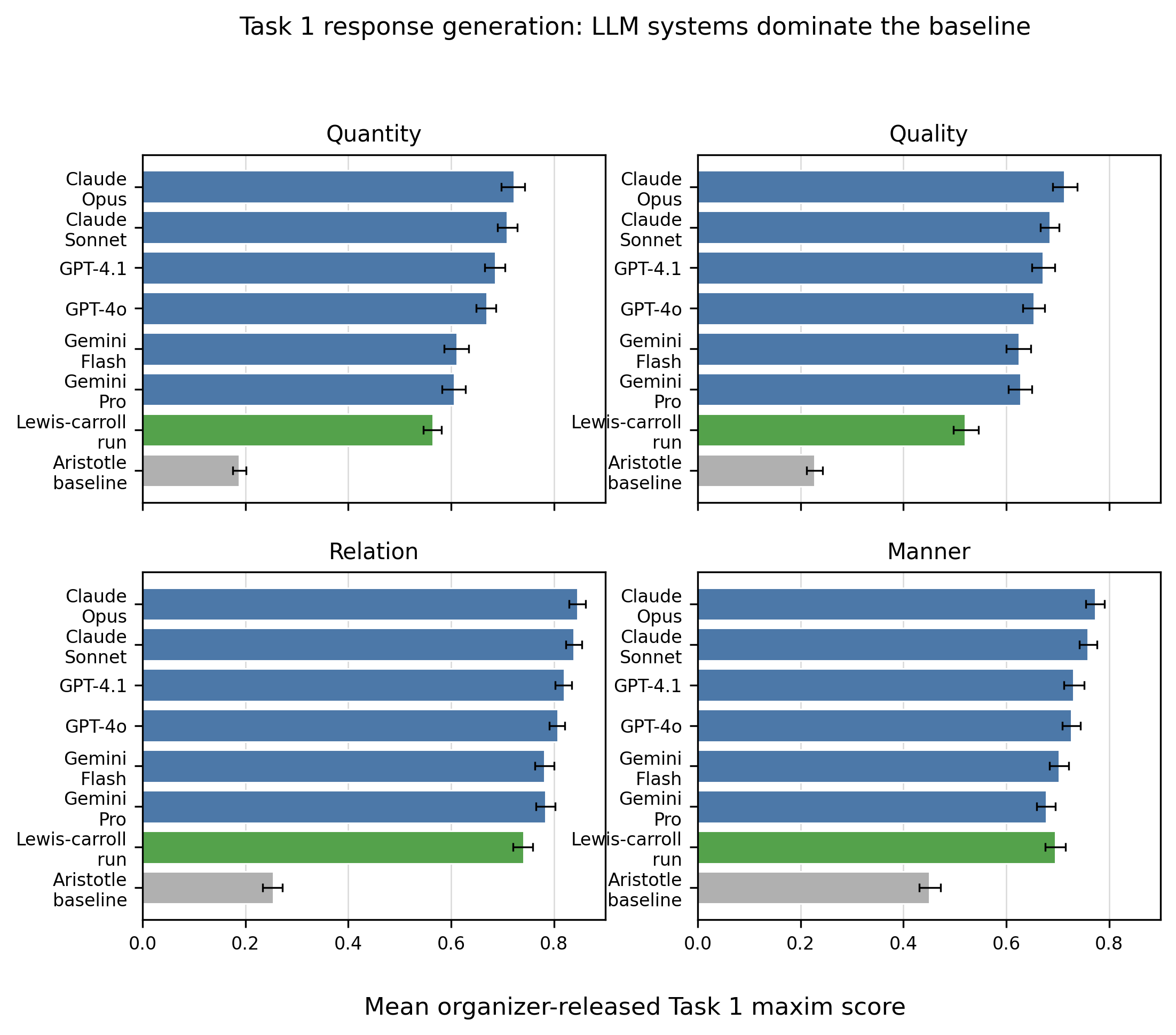}
\caption{
    Average subtask 1 maxim scores by response-generation system, averaged over released judge scores on the organizer-released corpus; this differs from the official Subtask 1 metric.
}
\label{fig:task1-system-scores}
\end{figure}

\subsection{Evaluator Consensus and Calibration}

\begin{figure}[tb]
\centering
\includegraphics[width=\textwidth]{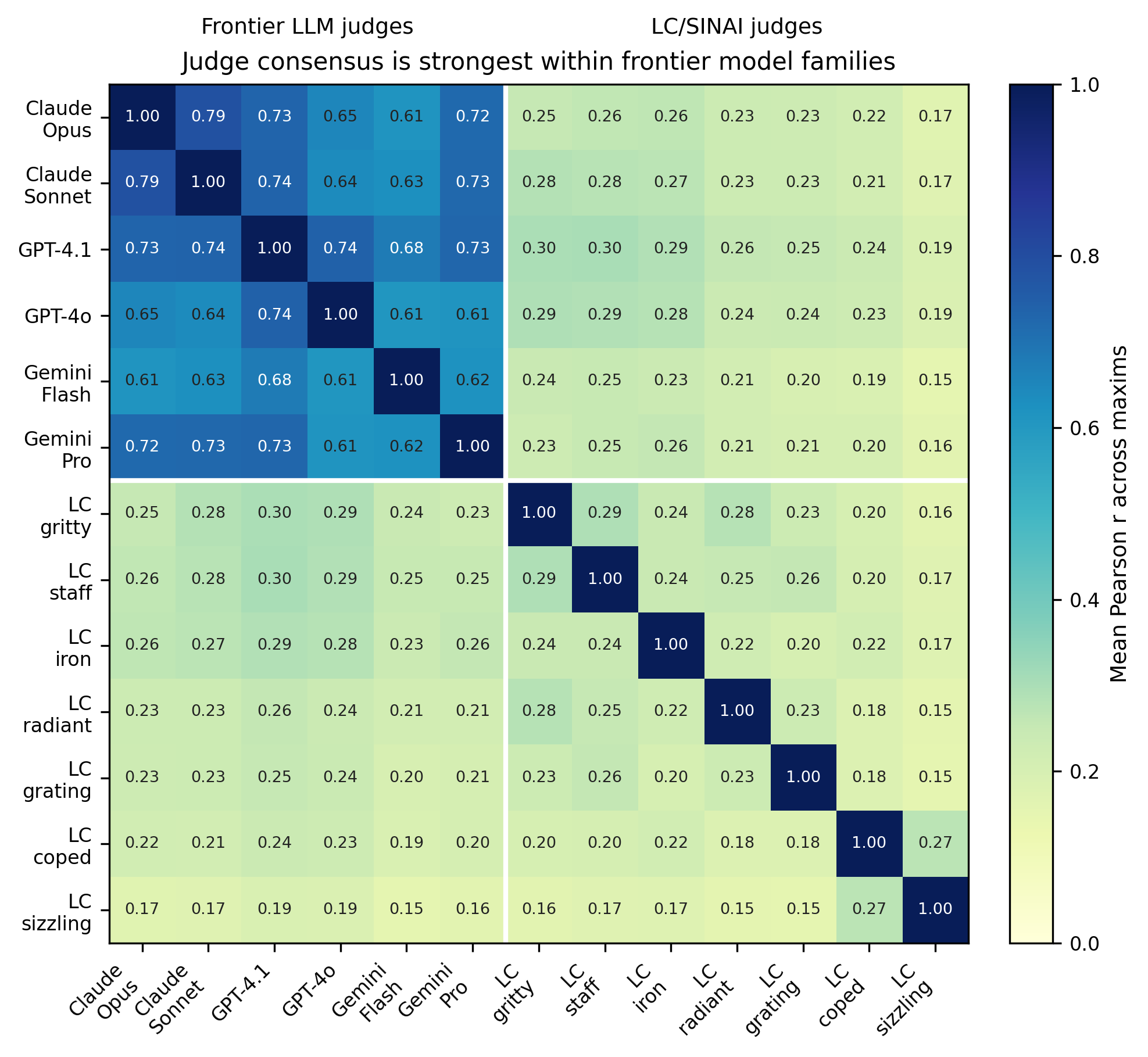}
\caption{Mean pairwise judge-score correlation across Gricean dimensions. Frontier LLM judges show strong agreement, especially within model families, while SINAI LLaMA-8B judges are less correlated with the frontier judge pool.}
\label{fig:judge-correlation-heatmap}
\end{figure}

Under the averaged judge scores in Figure~\ref{fig:task1-system-scores}, the DS@GT systems outperform the baseline, unlike under the official binary annotations in Table~\ref{tab:touche25_subtask1_results}.
Frontier LLM judges show substantial agreement during judging, but Figure~\ref{fig:judge-correlation-heatmap} shows that this agreement is structured.
Same-family judges are especially correlated.
Claude Opus and Claude Sonnet have an average correlation of 0.844 across Gricean dimensions, while GPT-4.1 and GPT-4o average 0.772.
Across the frontier judges, correlations are generally high for Quantity, Quality, and Relation, ranging from roughly 0.65 to 0.91.
This means that high agreement should not be treated as fully independent evidence of evaluator validity.
The strongest agreement is among model-family clusters, suggesting that consensus may reflect shared model behavior rather than independent evaluator convergence on response quality. 
However, the LLaMA-8B judges submitted by SINAI correlate much less with frontier judges, ranging between 0.17 and 0.27.
Additionally, they correlate weakly with each other, ranging between 0.15 and 0.29.
Even though these judges are weaker, they provide less redundant signal, and the distinction between judge strength and judge diversity is pivotal in the following calibration analyses.

After calibrating consensus among judges, we build a weakly-supervised label model to see how it might rank against the others in a post-competition setting.
The frontier-only independent label model was nearly indistinguishable from the raw average rankings, with Kendall's $\tau = 1.0$ across all four Gricean dimensions.
This suggests that when the voter pool is dominated by highly correlated frontier LLM judges, the label model recovers the same consensus already present in the raw average.
Therefore, weak-supervision did not substantially change the ranking when the voters agreed in similar ways.
Adding the SINAI judges moved the rankings more because these judges were less redundant than the frontier models.
With the expanded judge pool, Kendall's $\tau$ between calibrated and raw rankings dropped to 0.93 for Quantity, 0.86 for Quality, 0.93 for Relation, and 0.71 for Manner in Table~\ref{tab:weak-supervision}. 
This does not mean smaller judges were superior evaluators overall, only that they contributed less redundant signal than the frontier LLMs.

Additionally, we tested a dependency-aware version of the label model, where correlated voters such as same-family LLMs and length-related heuristics were explicitly linked.
This improved alignment with the official Subtask 2 F1, but only modestly.
Spearman correlation increased from 0.366 to 0.408 for Quantity, 0.224 to 0.286 for Quality, 0.500 to 0.580 for Relation, and 0.011 to 0.075 for Manner, suggesting that dependency-aware modeling discounts some redundancy in agreement, but it cannot build official-target signal when judges are highly correlated.
The gap between consensus and the official target is largest for Quality, where dependency-aware latent accuracy averages 0.687 while mean official Quality F1 is 0.244.
Overall, the weak-supervision experiments are useful for diagnosing consensus structure, but that consensus does not necessarily imply correctness.

\begin{table}[t]
\caption{Weak-supervision calibration summary. Kendall's $\tau$
compares calibrated system rankings with raw average rankings; Spearman's
$\rho$ compares latent judge-accuracy estimates with official Subtask 2 F1.}
\label{tab:weak-supervision}
\centering
\small
\setlength{\tabcolsep}{4pt}

\begin{tabular}{@{}lcccc@{}}
\toprule
\textbf{Analysis}
& \textbf{Quantity}
& \textbf{Quality}
& \textbf{Relation}
& \textbf{Manner} \\
\midrule

Expanded pool: $\tau$ vs.\ raw
& 0.930 & 0.860 & 0.930 & 0.710 \\

Dependency-aware: $\tau$ vs.\ raw
& 0.857 & 1.000 & 1.000 & 0.786 \\

Expanded independent pool: $\rho$ vs.\ F1
& 0.366 & 0.224 & 0.500 & 0.011 \\

Dependency-aware: $\rho$ vs.\ F1
& 0.408 & 0.286 & 0.580 & 0.075 \\

\bottomrule
\end{tabular}
\end{table}



\section{Discussion}
\label{sec:discussion}


The Quality dimension was the clearest failure mode in our analysis.
Quality is challenging in retrieval-augmented debate because it asks for more than a fluent, well-structured, relevant response.
A high-quality response should be factually correct, supported by retrieved evidence, and useful within the debate context.
In that sense, Quality overlaps with the other Gricean maxims while also requiring additional verification \cite{grice1975logic}.
This helps explain why the gap between evaluator consensus and official F1 was largest for Quality.
LLM judges can agree on a response that appears convincing, but that does not necessarily mean the response is actually supported by evidence. 

Subtask 2 ended up being one of the most interesting and valuable parts of the lab because Quality was so challenging to reliably rank.
This allowed us to look beyond which systems generated strong responses and ask how evaluator systems behaved.
The main limitation is that the Gricean maxims are broad evaluation targets.
They are useful as high-level categories, but they do not fully specify what a judge should check, especially for Quality.
Quality can include factuality, evidence support, reasoning validity, and argumentative usefulness.
If these are collapsed into a single score, judges may agree on surface plausibility without performing the same underlying evaluation.

\section{Future Work}

We aim to explore finer-grained control over the agent's behavior.
It would be beneficial to modify the agent's behavior to more accurately simulate diverse perspectives and personalities and to encourage participants to strategize in ways that capitalize on the strengths and exploit the weaknesses in discourse.
Behaviors can be implemented via controllable parameters that govern specific strategy types and explored by having several sets of behaviors, which, while not necessarily realistic, are easy to examine under controlled conditions.
Strategies could include always deflecting, always attacking, or always conceding.

Another area of future work is formulating a more robust evaluation framework.
Although we have a basic evaluation system, having more time to explore the results of the evaluation between different models would be interesting in the context of a leaderboard.
These results have to be grounded in human evaluation, but an ELO system could be used to rank models based on their performance in debates.

Future RAD labs should report evaluator reliability alongside leaderboard scores, including inter-judge agreement, score distributions, model-family effects, length sensitivity, and dimension-level calibration.
For Quality, future evaluations should include more explicit verification, such as separating factuality, evidence support, reasoning validity, and argumentative usefulness instead of collapsing them into one broad score.

\section{Conclusions}

Frontier LLM judges often agree strongly with one another, but this consensus is partly explained by their family and thus the underlying training corpus.
Consensus-based latent accuracy estimates do not reliably align with the official Subtask 2 F1, with the largest gap on the Quality maxim.
Therefore, the question is not only whether LLMs can debate effectively, but whether retrieval-augmented debate can be evaluated reliably through multi-LLM agreement on Gricean maxims \cite{grice1975logic}.
The accompanying source code for this paper is located at \url{https://github.com/dsgt-arc/touche-2025-rad} and \url{https://github.com/dsgt-arc/touche-2025-rad-analysis}.

\begin{credits}
\subsubsection{\ackname}
We thank the Data Science at Georgia Tech (DS@GT) CLEF competition group for their support. During the preparation of this work, the authors used Gemini and Claude for grammar and spelling checks, formatting assistance, and peer review simulation; the authors subsequently reviewed and edited the content and each take full responsibility for the publication's content.

\subsubsection{\discintname}
The authors have no competing interests to declare that are relevant to the content of this article.
\end{credits}

\bibliographystyle{style/splncs04}
\bibliography{main}
\end{document}